\documentclass{optica-article}

\journal{opticajournal} 

\articletype{Research Article}

\usepackage{lineno}

\begin{document}

\title{FPM-INR: Fourier ptychographic microscopy image stack reconstruction using implicit neural representations}

\author{Haowen Zhou,\authormark{1,\dag} Brandon Y. Feng,\authormark{2,\dag} Haiyun Guo,\authormark{3} Siyu (Steven) Lin,\authormark{1} Mingshu Liang,\authormark{1} Christopher A. Metzler,\authormark{2,*} and Changhuei Yang\authormark{1,*}}

\address{\authormark{1}Department of Electrical Engineering, California Institute of Technology, Pasadena, CA, 91125, USA\\
\authormark{2}Department of Computer Science, The University of Maryland, College Park, MD 20742, USA\\
\authormark{3}Department of Electrical and Computer Engineering, Rice University, Houston, TX, 77005, USA\\
\authormark{\dag} The authors contributed equally to this work.}
\email{\authormark{*}chyang@caltech.edu, metzler@umd.edu} 

\begin{abstract*} 
Image stacks provide invaluable 3D information in various biological and pathological imaging applications. 
Fourier ptychographic microscopy (FPM) enables reconstructing high-resolution, wide field-of-view image stacks without $z$-stack scanning, thus significantly accelerating image acquisition. 
However, existing FPM methods take tens of minutes to reconstruct and gigabytes of memory to store a high-resolution volumetric scene, impeding fast gigapixel-scale remote digital pathology.
While deep learning approaches have been explored to address this challenge, existing methods poorly generalize to novel datasets and can produce unreliable hallucinations.
This work presents FPM-INR, a compact and efficient framework that integrates physics-based optical models with implicit neural representations (INR) to represent and reconstruct FPM image stacks.
FPM-INR is agnostic to system design or sample types and does not require external training data. In our demonstrated experiments, FPM-INR substantially outperforms traditional FPM algorithms with up to a 25-fold increase in speed and an 80-fold reduction in memory usage for continuous image stack representations.

\end{abstract*}

\section{Introduction}
Computational microscopy models the forward propagation of a light field, from illumination and light-sample interaction to sensor measurement formation, and then computationally inverts this forward model to form an image.
This fusion of optics and algorithms allows computational microscopy to offer substantial advantages over traditional brightfield microscopy. Computational microscopy has improved microscope resolution~\cite{zheng_wide-field_2013}, imaging speed~\cite{chung_wide-field_2016}, cost~\cite{aidukas_low-cost_2019, bian_ptychographic_2020}, and field-of-view~\cite{jiang_high-throughput_2022}; has enabled quantitative phase retrieval~\cite{chen_3d_2016, baek_kramerskronig_2019, zuo_transport_2020, ling_high-throughput_2018}; and has unlocked new capabilities such as automatic aberration correction~\cite{Bian:13,ou_embedded_2014} and digital refocusing\cite{tippie_high-resolution_2011, liang_all--focus_2022}. Computational microscopy is now widely used in biological\cite{popescu_chapter_2008,baek_intensity-based_2021}, clinical\cite{wang_optical_2023}, and pathological imaging\cite{horstmeyer_digital_2015}; non-invasive surface inspection\cite{shen_non-iterative_2021, zhou_review_2022,wang_fourier_2023}; and aberration metrology\cite{memmolo_automatic_2011, ou_embedded_2014}. Fourier ptychographic microscopy (FPM), which enables wide field-of-view imaging, is one of the most successful and widely utilized computational microscopy techniques and has been extensively studied since 2013~\cite{zheng_wide-field_2013, ou_high_2015,zheng_concept_2021}. 

One of the most important features of FPM is its ability to correct for aberrations, notably defocus, {post-capture}.
Defocus aberration manifests when the region of interest within the specimen deviates from the front focal plane of the microscope objective lens. This deviation from the ideal focal point may arise from various factors, including the inclined disposition of the sample and sample unevenness across the region. With its digital refocusing capability, FPM can computationally reconstruct optical fields at distinct planes situated along the optical axis. Consequently, this functionality not only eliminates the need to perform physical re-scanning, but also facilitates sparse volumetric ($z$-stack) imaging. 
If the sample contents are distributed sparsely within the volume, then the sample can be approximated and reconstructed as a succession of 2D cross-sections~\cite{liang_all--focus_2022}.
This approximation is valid for a range of digital pathology slide analyses such as those from fine needle biopsy aspirates \cite{pmid40687,liang_all--focus_2022} and brain tumor biopsies \cite{conway1973stereotaxic,ostertag1980stereotactic}. 

Laser illumination allows FPMs to acquire all the measurements required to form a high-resolution wide field-of-view volume within a second~\cite{Chung:16}.
However, the computational demands of current FPM reconstruction algorithms remain a significant obstacle for high-throughput pathological imaging applications.
Existing FPM algorithms reconstruct each slice of a $z$-stack image independently, solving a time-consuming optimization problem for each slice. As a result, reconstructing a high-resolution $z$-stack can take tens of minutes on a Graphics Processing Unit (GPU) (Nvidia RTX A6000), which is impractically slow for interactive pathology applications.
Moreover, the $z$-stacks generated by existing FPM algorithms are high-dimensional data, leading to high storage and transmission costs.
This inhibits the broader integration of FPM into digital pathology~\cite{BANERJEE2020121} and collaborative diagnosis~\cite{lu_federated_2022,ogier_du_terrail_federated_2023}, where there is a growing need for remote diagnosis, inter-institutional data transfer, and compact and efficient data packaging.
An attempt has been made with a deep learning method to tackle such challenges \cite{bouchama_fourier_2023}, but it requires external training data and depends on system design and sample types (details in Section~\ref{sec:FPMrecon}).

In this work, we introduce a compact, computationally efficient, and physics-based framework for reconstructing and representing FPM image stacks, termed Fourier ptychographic microscopy with implicit neural representation (FPM-INR). FPM-INR combines implicit neural representations (INRs), efficient volume decomposition, GPU acceleration, and strategic optimization, to efficiently solve the FPM image stack reconstruction problem. 

The difference in data representations between conventional FPM and the proposed FPM-INR is particularly noteworthy. FPM generates a $z$-stack with the same architecture as a physical $z$-stack, i.e., a Cartesian volume of M $\times$ N $\times$ P voxels, where M and N represent the lateral pixel counts and P represents the z-gradation.  In contrast, FPM-INR encapsulates the physical $z$-stack data into a compact feature volume coupled with the weights of a small neural network. In essence, the pattern and sparsity of the sample are efficiently captured by the novel parameter space of FPM-INR.

FPM-INR leverages the known physics-based FPM forward model and is compatible with any FPM microscope without necessitating hardware modifications. In addition, it does not require any pre-training. In our demonstrated experiments, FPM-INR can reduce the reconstructed data volume by $80\times$, accelerate the reconstruction process by up to $25\times$, and generate image stacks with fewer artifacts. 
We outline and explain FPM-INR in Section~\ref{sec:Method}. Experiments in Section~\ref{sec:Results} validate our method, where we quantitatively compare the quality, time, and data storage performance of our method with the conventional FPM approach, and we demonstrate its applicability from a human blood smear sample to cytology imaging of thyroid gland lesions. Section~\ref{sec:Discussion} summarizes the key features and concepts of our method and discusses implications to broader applications of FPM.

\section{Related Work}\label{sec:Related}
\subsection{FPM Reconstruction}\label{sec:FPMrecon}
FPM processing is typically performed with a combination of alternating projection algorithm and embedded pupil function recovery algorithm \cite{zheng_wide-field_2013,ou_embedded_2014}. Some of the recent FPM developments center on improving reconstruction quality or adapting to challenging scenarios. To date, only a few of these developments have attempted to speed up the reconstruction process and/or alleviate the massive computation load in $z$-stack imaging. One proposed approach solves the FPM imaging problem through neural network modeling in a forward pass~\cite{jiang_solving_2018}. This method speeds up the FPM reconstruction by taking advantage of the GPU acceleration for 2D phase retrieval. However, adapting this method to $z$-stack imaging would simply include an additional loop to the reconstruction pipeline, which neither exploits the inherent anisotropic optical resolution nor reduces the data volume. 

Another type of attempt is through digital refocusing in a post-reconstruction manner. One proposed solution~\cite{claveau_digital_2020} is to digitally propagate the optical field after the FPM reconstruction to obtain focused images at different planes. If feasible, this would greatly simplify $z$-stack image generation. Unfortunately, this approach violates the physics principle of the FPM forward model and has been demonstrated to be problematic~\cite{zhou_analysis_2022}. 

Deep learning has been explored in the context of post-reconstruction digital refocusing, where a deep neural network is trained with supervised learning to learn a prior over $z$-slices \cite{bouchama_fourier_2023}.
This method can reduce the image stack data volume and quickly generate images of different slices, but deep-learning-based methods generally have several limitations, including (a) a strict requirement of a large dataset with defocus distance values; (b) a computationally intensive training process; (c) the susceptibility to generalization challenges under unseen sample categories; (d) the reliance on a particular system design that the model is trained under, including factors like illumination patterns, numerical apertures (NA) of the objective lenses, and camera and magnification settings; (e) the restriction to a set of discrete $z$-planes.
The constraints inherent to conventional deep learning methods pose significant issues for digital pathology applications, where even minor inaccuracies are unacceptable due to the critical nature of the context.

\begin{figure}[t!]
\centering\includegraphics[width=13cm]{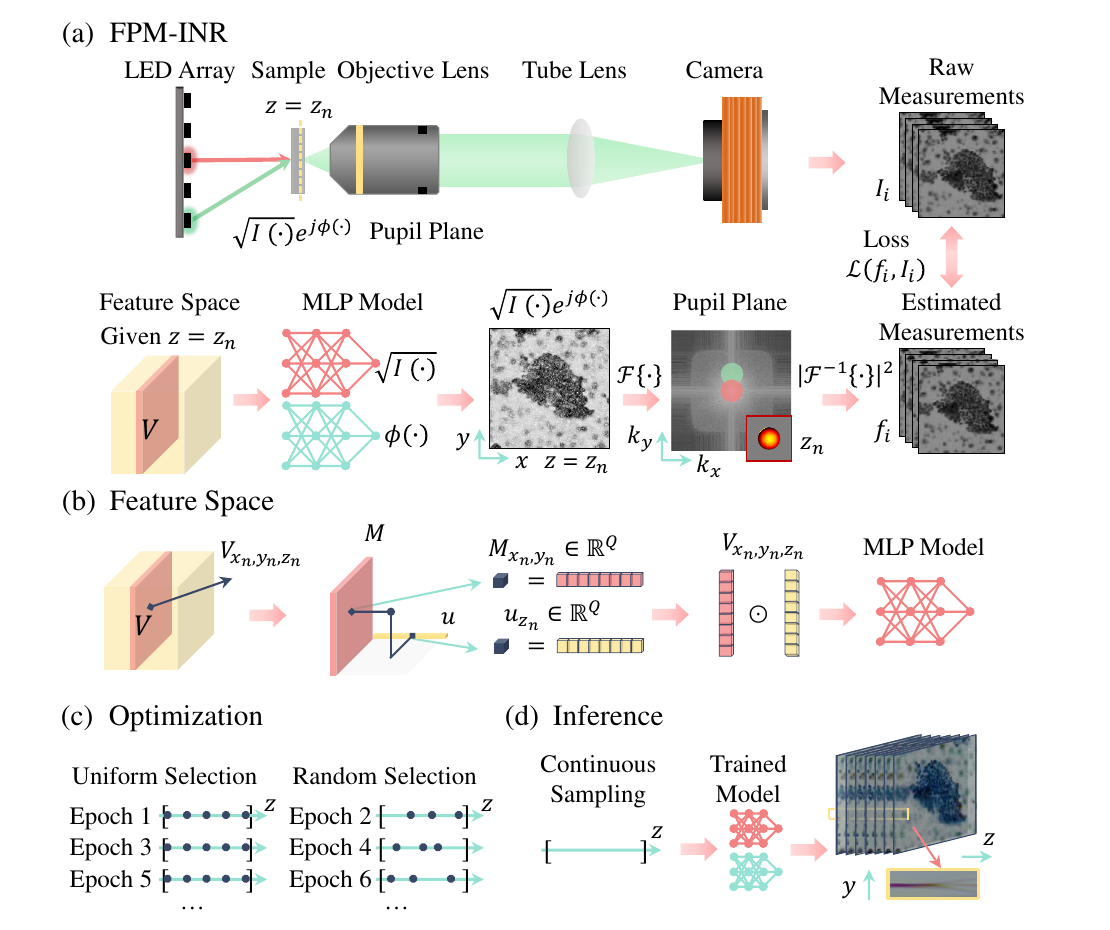}
\caption{(a) General framework of FPM-INR. FPM-INR starts at random initialization of the feature space volume. The multi-channel feature vector for each point is input to the MLP model. The output of the MLP model is an estimate of the value at the corresponding point in the high-resolution optical field. This estimated optical field represents the complex sample function. After MLP inference, the resulting high-resolution field goes through FPM's physics-based forward model related to the optical setup from illumination to camera.
The forward model outputs the estimated measurements, and the difference between the estimated and raw measurements is used to update the model weights and feature space parameters. $V$ represents feature space volume; $I$ and $\phi$ are intensity and phase; $\mathcal{F}$ is Fourier transform; $k_x, k_y$ are spatial frequency coordinates. (b) Feature space design. Instead of explicitly storing every 3D voxel in the feature space volume, we only learn a 2D feature plane $M$ and a 1D feature vector $u$. To obtain the feature vector $V_{x_n,y_n,z_n}$ for a point $(x_n, y_n, z_n)$, we project $(x_n, y_n)$ onto $M$ and $z_n$ onto $u$, sample feature vectors $M_{x_n, y_n}$ and $u_{z_n}$ with continuous bilinear interpolation, and compute the elementwise Hadamard product between $M$ and $u$. $Q$ is the number of feature channels. (c)  $z$-slices selection strategies. We select different image stacks over the optimization process. Each black dot denotes a sampled value on the $z$-axis. (d) Continuous inference. After training, FPM-INR supports continuous inference at arbitrary sampled values on the $z$-axis.}
\end{figure}

\subsection{Implicit Neural Representations}
The limitations of prior studies strongly indicate that a physics-based and fast FPM reconstruction technique with low data volume representation is highly desirable, but it is missing from the current state-of-the-art. We propose using implicit neural representation (INR) to address this gap. INR is a relatively new computational concept centered on mapping spatial coordinates to image pixel values with a multi-layer perceptron (MLP) model acting as a continuous mapping function \cite{sitzmann2019siren, park2019deepsdf, mildenhall2020nerf}. This concept has been instrumental in the recent advances of computer vision, computer graphics, and generative artificial intelligence \cite{mildenhall2020nerf, Pumarola_2021_CVPR, feng2021signet, sitzmann2019siren,park2019deepsdf,feng2022prif, feng2023neuws, chan2022efficient}.
However, few studies have applied INR in the context of computational microscopy.
A recent work~\cite{zhu_dnf_2022} used INR in lensless microscopic imaging to map 2D spatial coordinates to 2D amplitude and phase with an embedded forward model. A concurrent work~\cite{liu_recovery_2022} applied INR to intensity diffraction tomography to achieve a continuous recovery of a volumetric refractive index map; the method has been improved in a later work~\cite{xie_diner_2022} by adding a learnable hash encoding layer to speed up the convergence of the algorithm. 
These works employ the MLP model as an encoder-decoder functionality, which is computationally intensive. 
A more recent work~\cite{wang_local_2023} applied the MLP model as a decoder and trained convolutional neural networks as encoders to extract features from raw measurements. Their work achieved wrapping-free phase retrieval for 2D samples with fewer artifacts compared to conventional quantitative phase imaging techniques. 

\section{Method}\label{sec:Method}

\subsection{General Framework}
Our FPM-INR framework for image stack reconstruction is depicted in Fig. 1. The INR renders the high-resolution optical field from random initialization and is self-supervised by the FPM measurements through the physics forward model of FPM.

First, an FPM optical system is modeled mathematically from illumination to detection. The oblique LED illumination on the sample can be approximated by a plane wave. The plane wave modulated by the complex sample function $o(x,y;z)$ then is transferred to the pupil plane of the image system by an optical Fourier transform. At the pupil plane, the oblique angle illumination is converted to the lateral translations of the sample spectrum. By utilizing various angles of the illuminations both low and high spatial frequency components can be covered and captured. A set of raw measurements $I_i(x,y;z)$ associated with different illumination angles can be obtained by the tube lens performing an inverse Fourier transform. The forward model can be explicitly expressed as:
\begin{equation}
I_i(x,y;z) = |\mathcal{F}^{-1}\{O(k_x-k_{x_i},k_y-k_{y_i}) P(k_x,k_y;z)\} |^2
\label{eq1}
\end{equation}
where $I_i(x,y;z)$ is the measurement from $i^{th}$ LED illumination; $z$ indicates the defocus distance (from the sample to the front focal plane of the objective lens), which corresponds to the pre-defined quadratic defocus aberration added to the phase of the pupil function; $\mathcal{F}^{-1}$ is the inverse Fourier transform operator; $O(k_x-k_{x_i},k_y-k_{y_i})$ is the spectrum of the $o(x,y;z)$ from $i^{th}$ LED illumination; $P(k_x,k_y;z)$ is the pupil function; $k_x$ and $k_y$ are spatial frequency coordinates. 

For simplicity, we start introducing our framework with a 2D thin sample. Our FPM-INR framework tries to solve the problem by modeling the forward pass of FPM (Eq. \eqref{eq1}).  The mapping between the optical system and the physics-based forward model embedded in our framework is depicted in Fig. 1(a). The framework begins with the random initialization of the feature space volume. The feature vectors for each point are then taken as the input to two MLP models, each predicting the amplitude $\sqrt{I(\cdot)}$ and phase $\phi(\cdot)$ of a high-resolution complex field $\sqrt{I(\cdot)} \exp(j\phi(\cdot))$. This high-resolution complex field can be considered as an analog to the complex sample function. Illuminated by an oblique plane wave, this high-resolution complex field propagates through the objective lens and covers a part of the spectrum at the pupil plane as highlighted in the green circular region in Fig. 1(a). The corresponding spectrum then formulates an estimated measurement ($f_i$) through an inverse Fourier transform and a square function. This resembles the functionality of the tube lens and the camera in the optical system. The optimization objective minimizes the difference (smooth L1 loss) between captured raw measurements and estimated measurements. Subsequently, the weights of the MLP model and parameters ($M$ and $u$) of the feature space volume are updated through gradient descent. After iterating the above process till convergence, the high-resolution complex field is reconstructed. The $z$-dimension will be introduced in Section \ref{sec:opt}.

\subsection{Feature Space Design}

To model a volumetric sample, instead of explicitly storing each discrete 3D voxel with its complex value, we construct a feature volume $V$ (Fig. 1(b)), where each voxel stores a learnable $Q$-channel feature vector: $V_{x_n,y_n,z_n} \in \mathbb{R}^Q, n = 1,2,...N$.
The size of this feature volume may be smaller than the size of the digitized sample, and we can use bilinear interpolation to obtain the feature for any continuous spatial coordinate.
A compact MLP is trained to convert such a feature vector into the value at  $(x_n,y_n,z_n)$ in the field.

As the optical resolution for FPM is spatially anisotropic, with the lateral ($x$- and $y$-axis) resolutions higher than the axial ($z$-axis) resolution, we adopt a low-rank-decomposed representation of $V$ in practice. Specifically, we use a 1D vector $u$ to succinctly represent the variations along the $z$-axis, while maintaining a full-rank matrix $M$ to capture variations across $x$ and $y$.
Each location in $u$ and $M$ stores a $Q$-channel feature vector that can be updated during optimization.
To obtain the feature at a point $(x_n, y_n, z_n)$, we project $(x_n, y_n)$ onto $M$ and project $z_n$ onto $u$ to obtain feature vectors $M_{x_n, y_n}$ and $u_{z_n}$.
As illustrated in Fig. 1(b), the $Q$-channel feature vector at location $(x_n, y_n, z_n)$ in the 3D feature volume is the Hadamard product between the feature vectors $M_{x_n,y_n}$ and $u_{z_n}$:
\begin{equation}
V_{x_n,y_n,z_n} = M_{x_n,y_n} \odot u_{z_n}
\label{eq2}
\end{equation}
where $\odot$ denotes Hadamard product, and $M_{x_n,y_n}, u_{z_n} \in \mathbb{R}^{Q}$. Effectively, our design is equivalent to approximating a 3D volume through a tensor product between a 2D matrix and a 1D vector. This approach falls under tensor decomposition strategies~\cite{chen2022tensorf, fridovich2023k}  commonly used to parametrize a 3D volume represented by an INR, which can effectively enhance the INR's ability to represent signals while simultaneously reducing the number of required parameters.

Given a specific defocus distance $z=z_n$, we first obtain $V_{x_n, y_n, z_n}$, the Hadamard product between the feature vectors $u_{z_n}$ and $M_{x_n,y_n}$. With this $Q$-channel feature vector as input, the MLP model has $Q$ channels in its first layer. The MLP model consists of two non-linear layers following with ReLU activation function and a linear layer producing a final output value. To render a complex-valued high-resolution optical field, we use two real-valued MLPs with two feature space volumes, and these two MLPs produce the amplitude and phase parts of the complex output separately.  
The discretized pixel count of the feature plane $M$ in each feature channel is one-sixteenth of the amplitude or phase outputs. The gap between the pixel counts is addressed by the bilinear interpolation along the $x-$ and $y-$axis. 
Our neural representation is highly compact, comprising only a few thousand parameters, which facilitates the acceleration of reconstruction. 

\subsection{Optimization and Inference}\label{sec:opt}
To efficiently reconstruct the image stack, the key idea in our optimization strategy is to employ feature space interpolation and an alternating $z$-slice selection strategy. The optimization process requires selecting specific $z$ values denoting the defocus distance, and the defocus distances can be continuous values within a range of $[z_{min},z_{max}]$. 

The limits of the defocus distance range are determined by the FPM digital refocusing maximum capacity. The extended depth of field for FPM can be influenced by many practical factors --- including but not limited to the precision of LED position calibration, coherent area of the LED illumination, total synthetic numerical aperture, and the wavelength of the illumination light. As such, it is difficult to establish an analytical formula or even an empirical equation to quantify the digital refocusing capability of FPM. Therefore, the defocus distance range is generally assessed to be an empirical range of 3-6 times larger than the incoherent brightfield microscope depth of field, or sample thickness prior \cite{zheng_wide-field_2013, zuo_wide-field_2020}. 

To numerically change defocus distances, the conventional FPM method associates the arbitrary defocus distance with the defocus aberration in the Fourier domain. 
To fulfill this functionality in our method without unnecessarily learning infinitely many $z$-slices, we perform interpolation along the $z$-axis when sampling from the feature space. 
Within the digital refocusing capacity $[z_{min},z_{max}]$, we first determine a few $z$-planes with uniform separations and initialize their feature representation in $u$. Each feature vector stored in $u$ corresponds to a discretized point on the $z$-axis.
For any continuous $z$ value, we can linearly interpolate its two nearest discretized feature vectors on $u$ to obtain its feature vector.

As shown in Fig. 1(c), we select different $z$ values for optimization at different epochs. At each odd number epoch, $z_n$ values are selected uniformly corresponding to the discretization of $u$, and the resulting $u_{z_n}$ is multiplied with the lateral feature vector $M_{x_n,y_n}$. The product of these is then sent to the MLP model. At each even number epoch, $z_n$ values are selected randomly with the resulting $u_{z_n}$ obtained through linear interpolation. This selection strategy avoids naively sampling infinitely many $z$-planes for optimization and speeds up reconstruction. 

Once the weights of MLP and the feature volume parameters are optimized, these data are fixed and can be saved as storage data for the sample. 
During model inference (Fig. 1(d)), the feature space can be continuously sampled to generate the image stack. Our experiments reported in Section \ref{sec:Results} provide more context to this consideration.

\section{Results}\label{sec:Results}
\subsection{Proof of Concept}\label{sec:blood}
To validate our proposed method, we used a human blood smear slide (Carolina Biological Supply Company, Wright's stain) as the initial test target. We tilted the slide at a 4-degree angle to the optical axis of the microscope. An LED array (Adafruit 32$\times$32 LED matrix, 4 mm pitch) together with a 16-element LED ring was used for illumination. The illumination NA was matched with the objective lens' NA (Olympus PLN 10$\times$/0.25NA). In total, 68 LEDs were used for sequential illuminations. We imaged at a center wavelength of 522 nm. The sample was placed 74 mm from the LED panel. A monochromatic camera (Allied Vision Prosilica GT 6400) with a pixel pitch of 3.45 microns was used. All these components were installed and customized on an Olympus IX51 inverted microscope body. 

For comparison, we captured brightfield images in the same setup with all LEDs lit. To avoid non-uniform illumination patterns, a piece of lens wiper (Kimtech Science) was placed between the LED array and the sample to help scatter the illumination. The image stack captured under the incoherent illumination was taken as the ground truth for our image stack from $z=-20$ $\mu m$ to $z=20$ $\mu m$ with a step of $0.25 $ $\mu m$ (161 layers in the $z$-stack). Fig. 2(a) presents some brightfield microscope images.

Conventional FPM reconstruction algorithms have different variants \cite{yeh_experimental_2015}. The sequential gradient descent algorithm \cite{zheng_wide-field_2013,ou_embedded_2014} is chosen for our comparison purpose as it is generally considered to be faster than the second-order methods (sequential Gauss-Netwon algorithm) \cite{tian_multiplexed_2014} and the convex-base method (PhaseLift)\cite{horstmeyer_solving_2015}. For simplicity, we will refer to the sequential gradient descent algorithm as the "FPM algorithm" in the following text. To minimize aberration influence (except defocus aberration) on reconstruction quality and convergence speed, the central field of view of the camera was selected as the region of interest with 1024$\times$1024 pixels. To make a fair comparison, GPU parallel computing was also implemented for the FPM algorithm. To guarantee consistent good convergence, we ran 25 iterations of the FPM algorithm for each $z$-plane. The FPM reconstructed images are shown in Fig. 2(b). The full stack image reconstruction is presented in Supplementary Video S1. 

As introduced in Section \ref{sec:opt}, FPM-INR employs a $z$-plane selection strategy. Here, $z$-planes with a uniform separation of 5 microns were selected as candidates for odd-number epochs, while three $z$-planes were randomly chosen for optimization at even-number epochs. In total, a number of 15 epochs were completed to establish good convergence. The Adam optimizer \cite{kingma2014adam} was used with a learning rate of $10^{-3}$ and a learning scheduler with a 10 times learning rate decay for every 6 epochs. The related images are shown in Fig. 2(c). Due to the tilted sample geometry (Fig. 2(d)), the sample content was focused on a continuum of $z$-planes, providing a good example to validate the feasibility of our method, with sample information distributed in every slice. The full stack image reconstruction from FPM-INR is also presented in Supplementary Video S1. 

\begin{figure}[ht!]
\centering\includegraphics[width=13cm]{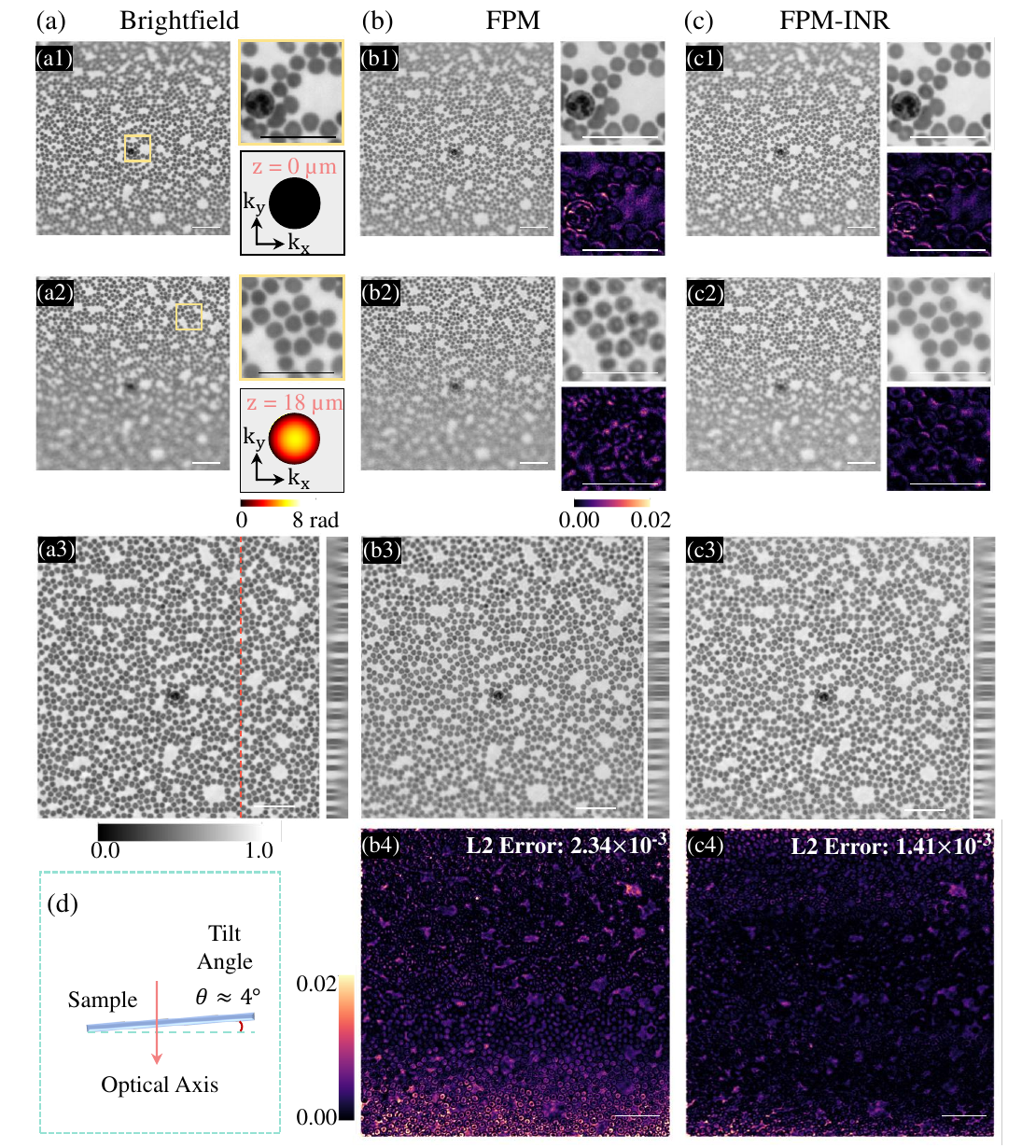}
\caption{Human blood smear image stacks from (a) brightfield microscope, (b) FPM, (c) FPM-INR. (a1,b1,c1) images at $z=0$ $\mu m$; (a2,b2,c2) at $z=18$ $\mu m$. The related zoom-in images, pupil phase, and L2 error maps are in the insets. (a3,b3,c3) all-in-focus images of all three methods. The red dashed line indicates the $yz$ cross section of the image stack. (b4,c4) the L2 error maps of FPM and FPM-INR, respectively. The scale bars are 50 $\mu m$. (d) A diagram of the sample geometry.}
\end{figure}

To evaluate our reconstructed image stack quality, both visual inspection and quantitative error metrics were applied. In general, FPM and FPM-INR can obtain similar image stack quality. From Figs. 2(a1,b1,c1), the images at $z=0$ $\mu m$ plane showed consistent quality for the white blood cell and red blood cells. In addition, the L2 error maps were computed by comparing FPM and FPM-INR images with brightfield measurement. The error maps and metrics indicated that our FPM-INR algorithm performed slightly better than the FPM algorithm. Another example for images at $z=18$ $\mu m$ led to the same conclusion. Additionally, the FPM-INR image had fewer artifacts than the FPM result compared with the ground truth image via a visual inspection. To further establish a quantitative analysis for reconstruction quality, the L2 error and error map were calculated over the all-in-focus images over the image stack. The all-in-focus images were constructed by using the normal variance method in Refs. \cite{bian_autofocusing_2020, liang_all--focus_2022}. FPM-INR (L2 error: $1.41\times 10^{-3}$) still gave better image quality than the FPM algorithm (L2 error: $2.34\times 10^{-3}$). Although our goal is not to boost the image stack reconstruction quality, we did observe that the FPM-INR algorithm reduces artifacts, especially at large defocus distances. 

To benchmark the compression ratio and time performances of FPM-INR v.s. conventional FPM, the same set of data was used on the same GPU device (Nvidia RTX A6000). The data volume size generated by the FPM was presented in Fig. 3(a). The high-resolution image stack had a size of 2048 pixels along lateral axes, and 161 $z$-slices along $z$-axis. In total, this data volume had 644 megapixels with 4 bytes for each pixel. This adds up to 2576 MB for the human blood smear sample. In contrast, FPM-INR only needs to save the feature space parameters and model weights (Fig. 3(a)). The feature plane $M$ had 512$\times$512 pixels covering the $xy$ plane, each storing a feature vector of $Q=32$ channels. The feature representation $u$ along the $z$-axis is uniformly discretized by 5 (number of pre-defined $z$-planes), each storing a feature vector of $Q=32$ channels. Interpolation is used to enable continuous sampling on $M$ and $u$. The feature parameters in total took up 32 MB in storage. The MLP model consisted of two non-linear layers and one linear layer with 32 neurons and 1 bias node. The number of weights can be calculated as $(32+1)\times 32 \times 2 + (32+1)\times 1 = 2145$, which is equivalent to 8.4 KB. Therefore, the total storage needed for FPM-INR was about 32 MB. The compression ratio, defined by FPM data volume over FPM-INR storage volume, achieved a factor of 80.5. The above calculations are done for amplitude images; to include phase images, the data volume and data storage size will be doubled for both FPM and FPM-INR.

We further examine the performance of FPM-INR and FPM at various patch sizes. Commonly, conventional FPM algorithms reconstructed square patches with sizes of $2^7$, $2^8$, $2^9$, and $2^{10}$ pixels along each lateral dimension. If the patch size is too small, the reconstruction may suffer from the lateral shift effect from oblique illumination at a large defocus plane (see Supplementary document). If the patch size is too large, the region of interest may exceed the coherent area of illumination which can be roughly estimated by the Van-Zernike-Cittert theorem \cite{konda_fourier_2020}. This coherent area is not a hard limit, but it would violate the coherent FPM forward model gradually. In our evaluation, the patch sizes were chosen considering that the fast Fourier transform algorithm prefers the image dimension to be of powers of two.

As shown in Fig. 3(b), the FPM-INR algorithm significantly outperformed the FPM algorithm in computational time with $9.8\times$, $11.8\times$, $7.5\times$, and $5.3\times$ increase for patch sizes of $2^7$, $2^8$, $2^9$, and $2^{10}$, on the same GPU device. This was confirmed across five experiments, as depicted by the error bars in Fig. 3(b). In addition, a compression ratio of about 80 times can be consistently achieved across different patch sizes, as indicated by the circle area in Fig. 3(b). The inference speed was approximately 460 MB/s on Nvidia RTX A6000 GPU for reference. This model inference time can be negligible in practice and will be further reduced with the rapid advancement of GPU devices. 

\begin{figure}[t]
\centering\includegraphics[width=13cm]{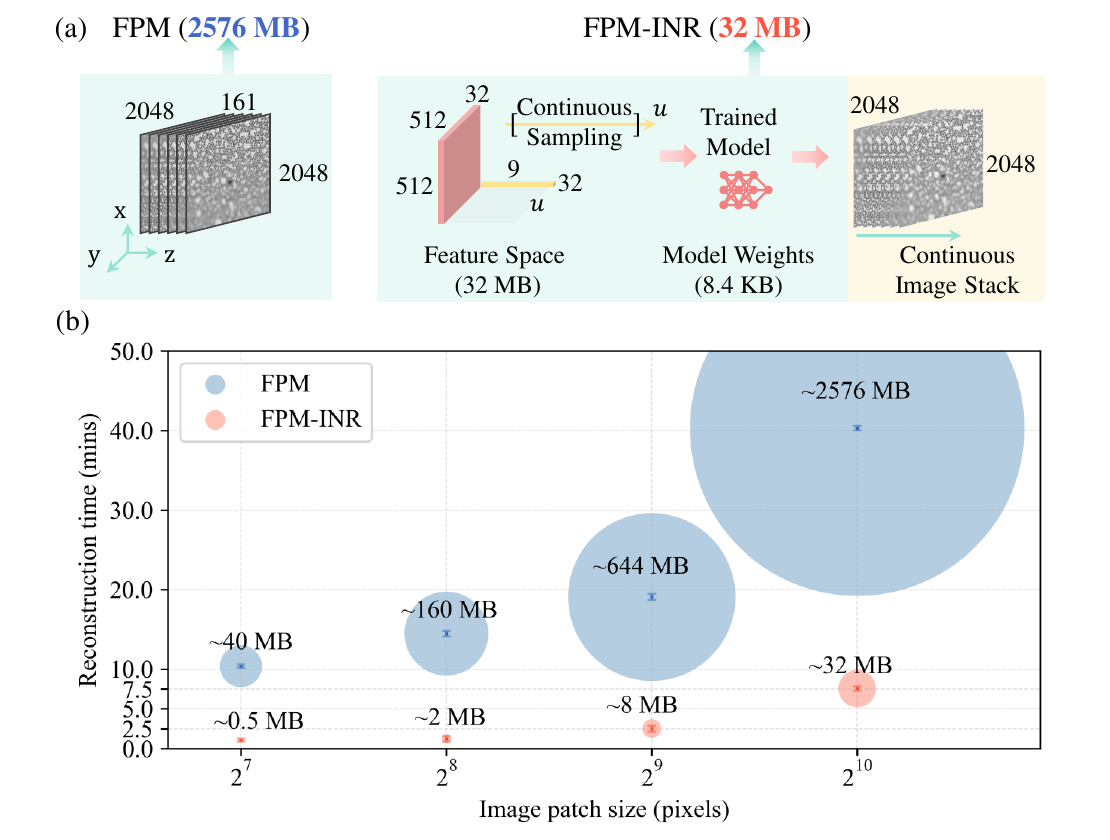}
\caption{(a) Data storage size by FPM and FPM-INR. The numbers are shown in pixels to demonstrate tensor or image sizes. Every pixel is 4 bytes in single-precision floating-point format. (b) Performance comparison between conventional FPM and FPM-INR on computational time and data storage size (the amplitude image stack only) across different patch sizes. The error bar indicates one standard deviation over five experiments. The circle area size linearly relates to the data size for storage. Compared to the FPM algorithm, FPM-INR is up to 12 times faster and reduces the data storage size by 80 times for the blood smear sample.}
\end{figure}

\subsection{Application to Digital Pathology}

Digital pathology is a growing application in clinical diagnosis and disease analysis. Cytology, also known as cytopathology, is a branch of diagnostic pathology that studies whole cells from bodily tissues and fluids. Our FPM-INR algorithm can further facilitate FPM digital pathology applications in these fields. Here we report a demonstration experiment where FPM-INR was used on a cytology specimen collected through thyroid fine needle aspiration. A fine needle aspiration biopsy Papanicolaou smear (pap smear) of papillary thyroid carcinoma was imaged by our system. Part of the data was obtained from Ref. \cite{liang_all--focus_2022}. The sample has a thickness of about $30$ $\mu m$ (from $-10$ $\mu m$ to $20$ $\mu m$) and cell aggregations at different heights. In clinical diagnosis, pathologists need to evaluate cellular structural information and color staining contrast over the whole sample volume. Regular brightfield microscope takes a long time to scan the sample for discrete $z$-slices and it results in a huge data volume. This hinders efficient data collaboration and quick pathological analysis. FPM relieves the burden from the massive scanning duty but still suffers from a long reconstruction time and a tremendous data size. The proposed FPM-INR framework can substantially solve the current dilemma. 

The sample was imaged by a 20$\times$/0.40NA objective lens with matched illumination NA using 145 LEDs, and the distance between the LED panel to the sample was 66 mm. A CCD camera (ON Semi KAI-29050, 5.5 $\mu m$ pixel pitch) was used to capture raw measurements. Similar to Section \ref{sec:blood}, the FPM algorithm was optimized for 121 z-slices, and FPM-INR was also implemented with the same set of hyperparameters: including the learning rate and scheduler, parameter initialization strategy, and the number of epochs. The differences were that in this case, the number of feature channels $Q$ was set to be 24, six planes were uniformly selected in the odd epochs, and three planes were randomly selected in the even epochs. The image stack reconstructions of FPM and FPM-INR are presented in part in Fig. 4. The full image stack reconstructions are presented in Supplementary Video S2.

In terms of the storage memory requirement, FPM-INR retains similar compression performance as in Section \ref{sec:blood}, achieving a data compression ratio of 80.5 on the thyroid gland lesion data across different patches. 
Using the same GPU,  FPM-INR is $24.7\times$, $17.9\times$, $7.3\times$, and $5.0\times$ faster than the FPM algorithm at patch sizes of $2^7$, $2^8$, $2^9$, and $2^{10}$.

\begin{figure}[ht!]
\centering\includegraphics[width=13cm]{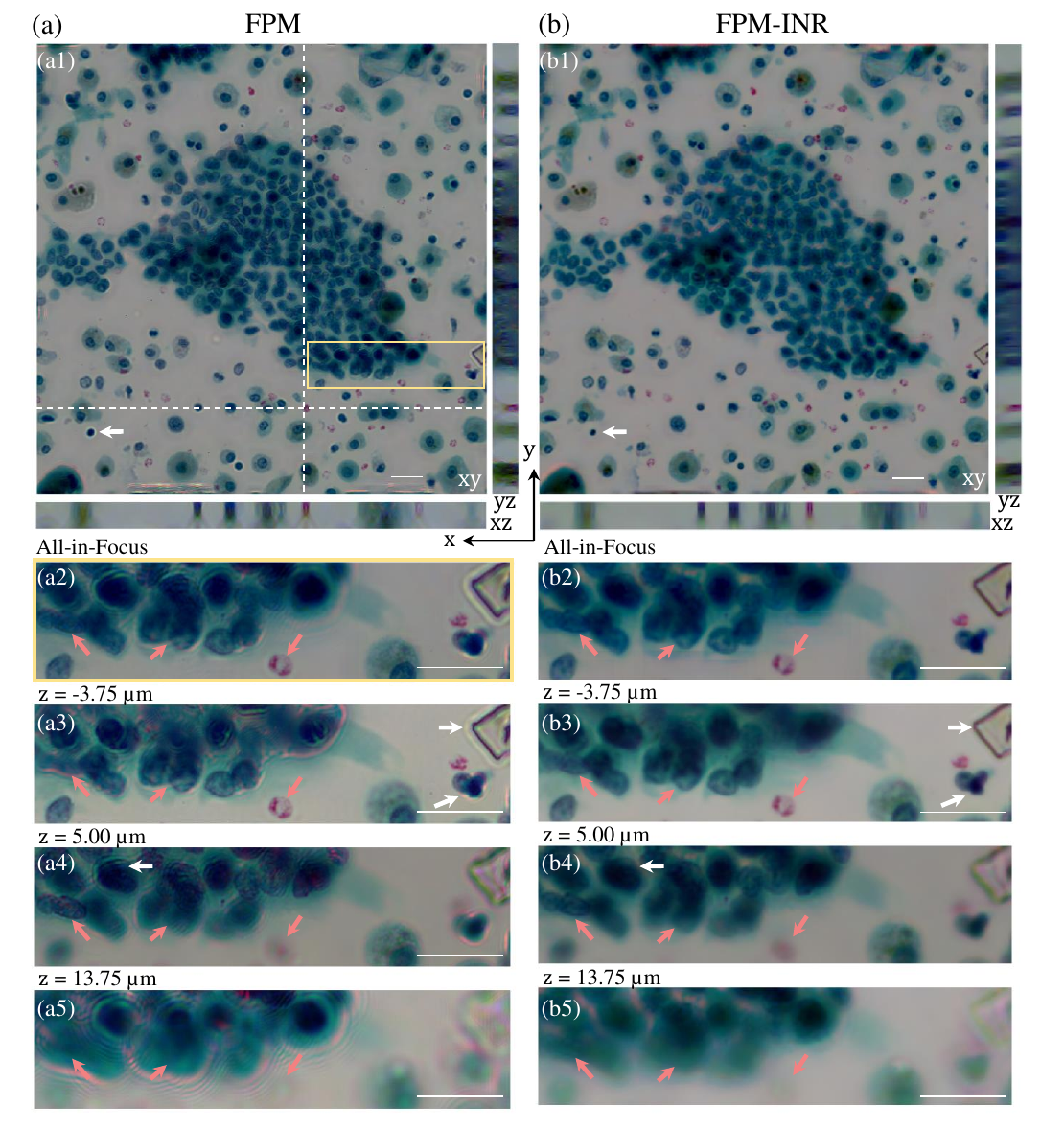}
\caption{Thyroid gland lesion pap smear images reconstructed by (a) FPM, (b) FPM-INR. (a1,b1) All-in-focus images with images at $xz$-plane and $yz$-plane along the dashed white lines. (a2,b2) Zoom-in images of the yellow box in (a1,b1). (a3-a5,b3-b5) Images at different $z$-planes. The red arrows point out the cell structure at different $z$-planes. The white arrows show the artifacts in FPM images, while FPM-INR is artifact-free. The scale bars are 20 $\mu m$.}
\end{figure}

Figures (a1,b1) present the all-in-focus images reconstructed by FPM and FPM-INR, respectively. The white dashed lines are associated with $xz$-plane and $yz$-plane sub-figures. The sub-figures along the $z$-axis demonstrate that FPM-INR digital refocusing quality is slightly better than the FPM algorithm. Figures (a2,b2) are the zoom-in section of the yellow box. The red arrows in Figures (a3-a5,b3-b5) are examples of cells focused at various depths. The white arrows in Fig. 4 point out the artifacts in the FPM images, while the FPM-INR does not have any such artifacts in the corresponding regions. This observation is consistent with the experiment using the human blood smear slide in Section \ref{sec:blood}. Additional experiments were provided in the supplementary document.

\section{Discussion}\label{sec:Discussion}

The central challenges in high-throughput, high-resolution pathological imaging using FPM lie in the computational, storage, and bandwidth demands associated with reconstructing and transferring $z$-stacks. 
While deep learning methods, in principle, could address these challenges, existing approaches generalize poorly to new data and can produce hallucinations that violate physical constraints. 
In this study, we sidestep these issues by introducing FPM-INR, a compact, fast, and physics-informed FPM image stack reconstruction framework. In our demonstrated experiments with validation data including human blood smear and thyroid gland lesion pap smear specimens,
the FPM-INR framework speeds up FPM reconstruction by up to $25\times$ and compresses FPM $z$-stack data by $80\times$. 
Importantly, the image stack quality is also enhanced both qualitatively and quantitatively with fewer artifacts than the conventional FPM algorithm.

While the FPM-INR framework draws inspiration from research on neural networks and deep learning, FPM-INR is physics-based, fully respects the physical model underlying the FPM measurement process, and only changes how we represent the $z$-stack data. FPM-INR does not merely treat the neural network as a black-box predictor, but rather leverages neural network's unique strengths in learning useful features and non-linear interpolation strategies based on the gradient-based feedback from data. 

Moreover, unlike deep learning methods which often require pre-training on external datasets with specific discrete $z$-planes,  FPM-INR is broadly adaptable to any FPM setup, regardless of the hardware specifics like objective lens, LED numbers, camera pixel pitch, or image patch size. FPM-INR sidesteps the generalization issues that often plague purely data-driven deep learning approaches, especially in critical applications like healthcare.


The key innovations behind FPM-INR hold numerous advantages:
(a) The physics-based pipeline using INR significantly improves upon conventional methods, while avoiding artifacts and generalization issues commonly associated with deep learning methods.
(b) The proposed method moves away from operating solely on the domain of physical space, which often involves anisotropic optical resolution, to operating on a feature space with efficient representation. This new paradigm allows for complex, high-resolution signals in the physical domain to be efficiently represented and recovered in the feature space, in conjunction with a physics-based inference process involving a compact neural network.
(c) INR enables continuous representations compactly and efficiently, which is leveraged by FPM-INR to offer high-resolution sample visualization, and can further enable more streamlined pipelines for downstream tasks.

\begin{backmatter}
\bmsection{Funding}
H.Z., S.L., M.L., and C.Y. would like to thank the Heritage Research Institute for the Advancement of Medicine and Science at Caltech (HMRI-15-09-01). B.Y.F. and C.A.M. were supported in part by the AFOSR Young Investigator Program Award no. FA9550-22-1-0208.


\bmsection{Disclosure}
The authors declare no conflict of interest.

\bmsection{Data Availability Statement}
Code will be available through \href{https://github.com/hwzhou2020/FPM_INR}{GitHub}. Data is available at \href{https://doi.org/10.22002/7aer7-qhf77}{CaltechData}.

\bmsection{Supplemental Document}
See the supplement document and videos for supporting content and additional discussion.

\end{backmatter}


\bibliography{FPM-INR}

\begin{thebibliography}{10}
\newcommand{\enquote}[1]{``#1''}

\bibitem{zheng_wide-field_2013}
G.~Zheng, R.~Horstmeyer, and C.~Yang, \enquote{Wide-field, high-resolution {Fourier} ptychographic microscopy,} {\protect\JournalTitle{Nature Photonics}} \textbf{7}, 739--745 (2013).

\bibitem{chung_wide-field_2016}
J.~Chung, H.~Lu, X.~Ou, H.~Zhou, and C.~Yang, \enquote{Wide-field {Fourier} ptychographic microscopy using laser illumination source,} {\protect\JournalTitle{Biomedical Optics Express}} \textbf{7}, 4787 (2016).

\bibitem{aidukas_low-cost_2019}
T.~Aidukas, R.~Eckert, A.~R. Harvey, L.~Waller, and P.~C. Konda, \enquote{Low-cost, sub-micron resolution, wide-field computational microscopy using opensource hardware,} {\protect\JournalTitle{Scientific Reports}} \textbf{9}, 7457 (2019).

\bibitem{bian_ptychographic_2020}
Z.~Bian, S.~Jiang, P.~Song, H.~Zhang, P.~Hoveida, K.~Hoshino, and G.~Zheng, \enquote{Ptychographic modulation engine: a low-cost {DIY} microscope add-on for coherent super-resolution imaging,} {\protect\JournalTitle{Journal of Physics D: Applied Physics}} \textbf{53}, 014005 (2020).

\bibitem{jiang_high-throughput_2022}
S.~Jiang, C.~Guo, P.~Song, T.~Wang, R.~Wang, T.~Zhang, Q.~Wu, R.~Pandey, and G.~Zheng, \enquote{High-throughput digital pathology \textit{via} a handheld, multiplexed, and {AI}-powered ptychographic whole slide scanner,} {\protect\JournalTitle{Lab on a Chip}} \textbf{22}, 2657--2670 (2022).

\bibitem{chen_3d_2016}
M.~Chen, L.~Tian, and L.~Waller, \enquote{{3D} differential phase contrast microscopy,} {\protect\JournalTitle{Biomedical Optics Express}} \textbf{7}, 3940 (2016).

\bibitem{baek_kramerskronig_2019}
Y.~Baek, K.~Lee, S.~Shin, and Y.~Park, \enquote{Kramers–{Kronig} holographic imaging for high-space-bandwidth product,} {\protect\JournalTitle{Optica}} \textbf{6}, 45 (2019).

\bibitem{zuo_transport_2020}
C.~Zuo, J.~Li, J.~Sun, Y.~Fan, J.~Zhang, L.~Lu, R.~Zhang, B.~Wang, L.~Huang, and Q.~Chen, \enquote{Transport of intensity equation: a tutorial,} {\protect\JournalTitle{Optics and Lasers in Engineering}} \textbf{135}, 106187 (2020).

\bibitem{ling_high-throughput_2018}
R.~Ling, W.~Tahir, H.-Y. Lin, H.~Lee, and L.~Tian, \enquote{High-throughput intensity diffraction tomography with a computational microscope,} {\protect\JournalTitle{Biomedical Optics Express}} \textbf{9}, 2130 (2018).

\bibitem{Bian:13}
Z.~Bian, S.~Dong, and G.~Zheng, \enquote{Adaptive system correction for robust fourier ptychographic imaging,} {\protect\JournalTitle{Opt. Express}} \textbf{21}, 32400--32410 (2013).

\bibitem{ou_embedded_2014}
X.~Ou, G.~Zheng, and C.~Yang, \enquote{Embedded pupil function recovery for {Fourier} ptychographic microscopy,} {\protect\JournalTitle{Optics Express}} \textbf{22}, 4960 (2014).

\bibitem{tippie_high-resolution_2011}
A.~E. Tippie, A.~Kumar, and J.~R. Fienup, \enquote{High-resolution synthetic-aperture digital holography with digital phase and pupil correction,} {\protect\JournalTitle{Optics Express}} \textbf{19}, 12027 (2011).

\bibitem{liang_all--focus_2022}
M.~Liang, C.~Bernadt, S.~B.~J. Wong, C.~Choi, R.~Cote, and C.~Yang, \enquote{All-in-focus fine needle aspiration biopsy imaging based on {Fourier} ptychographic microscopy,} {\protect\JournalTitle{Journal of Pathology Informatics}} \textbf{13}, 100119 (2022).

\bibitem{popescu_chapter_2008}
G.~Popescu, \enquote{Chapter 5 {Quantitative} {Phase} {Imaging} of {Nanoscale} {Cell} {Structure} and {Dynamics},} in \emph{Methods in {Cell} {Biology},}  vol.~90 (Elsevier, 2008), pp. 87--115.

\bibitem{baek_intensity-based_2021}
Y.~Baek and Y.~Park, \enquote{Intensity-based holographic imaging via space-domain {Kramers}–{Kronig} relations,} {\protect\JournalTitle{Nature Photonics}} \textbf{15}, 354--360 (2021).

\bibitem{wang_optical_2023}
T.~Wang, S.~Jiang, P.~Song, R.~Wang, L.~Yang, T.~Zhang, and G.~Zheng, \enquote{Optical ptychography for biomedical imaging: recent progress and future directions [{Invited}],} {\protect\JournalTitle{Biomedical Optics Express}} \textbf{14}, 489 (2023).

\bibitem{horstmeyer_digital_2015}
R.~Horstmeyer, X.~Ou, G.~Zheng, P.~Willems, and C.~Yang, \enquote{Digital pathology with {Fourier} ptychography,} {\protect\JournalTitle{Computerized Medical Imaging and Graphics}} \textbf{42}, 38--43 (2015).

\bibitem{shen_non-iterative_2021}
C.~Shen, M.~Liang, A.~Pan, and C.~Yang, \enquote{Non-iterative complex wave-field reconstruction based on {Kramers}–{Kronig} relations,} {\protect\JournalTitle{Photonics Research}} \textbf{9}, 1003 (2021).

\bibitem{zhou_review_2022}
H.~Zhou, M.~M.~R. Hussain, and P.~P. Banerjee, \enquote{A review of the dual-wavelength technique for phase imaging and {3D} topography,} {\protect\JournalTitle{Light: Advanced Manufacturing}} \textbf{3}, 1 (2022).

\bibitem{wang_fourier_2023}
H.~Wang, J.~Zhu, J.~Sung, G.~Hu, J.~Greene, Y.~Li, S.~Park, W.~Kim, M.~Lee, Y.~Yang, and L.~Tian, \enquote{Fourier ptychographic topography,} {\protect\JournalTitle{Optics Express}} \textbf{31}, 11007 (2023).

\bibitem{memmolo_automatic_2011}
P.~Memmolo, C.~Distante, M.~Paturzo, A.~Finizio, P.~Ferraro, and B.~Javidi, \enquote{Automatic focusing in digital holography and its application to stretched holograms,} {\protect\JournalTitle{Optics Letters}} \textbf{36}, 1945 (2011).

\bibitem{ou_high_2015}
X.~Ou, R.~Horstmeyer, G.~Zheng, and C.~Yang, \enquote{High numerical aperture {Fourier} ptychography: principle, implementation and characterization,} {\protect\JournalTitle{Optics Express}} \textbf{23}, 3472 (2015).

\bibitem{zheng_concept_2021}
G.~Zheng, C.~Shen, S.~Jiang, P.~Song, and C.~Yang, \enquote{Concept, implementations and applications of {Fourier} ptychography,} {\protect\JournalTitle{Nature Reviews Physics}} \textbf{3}, 207--223 (2021).

\bibitem{pmid40687}
T.~S. Kline, L.~P. Joshi, and H.~S. Neal, \enquote{{{F}ine-needle aspiration of the breast: diagnoses and pitfalls. {A} review of 3545 cases},} {\protect\JournalTitle{Cancer}} \textbf{44}, 1458--1464 (1979).

\bibitem{conway1973stereotaxic}
L.~W. Conway, \enquote{Stereotaxic diagnosis and treatment of intracranial tumors including an initial experience with cryosurgery for pinealomas,} {\protect\JournalTitle{Journal of Neurosurgery}} \textbf{38}, 453--460 (1973).

\bibitem{ostertag1980stereotactic}
C.~Ostertag, H.~Mennel, and M.~Kiessling, \enquote{Stereotactic biopsy of brain tumors.} {\protect\JournalTitle{Surgical neurology}} \textbf{14}, 275--283 (1980).

\bibitem{Chung:16}
J.~Chung, H.~Lu, X.~Ou, H.~Zhou, and C.~Yang, \enquote{Wide-field fourier ptychographic microscopy using laser illumination source,} {\protect\JournalTitle{Biomed. Opt. Express}} \textbf{7}, 4787--4802 (2016).

\bibitem{BANERJEE2020121}
A.~Banerjee, C.~Chakraborty, A.~Kumar, and D.~Biswas, \enquote{Chapter 5 - emerging trends in iot and big data analytics for biomedical and health care technologies,} in \emph{Handbook of Data Science Approaches for Biomedical Engineering,}  V.~E. Balas, V.~K. Solanki, R.~Kumar, and M.~Khari, eds. (Academic Press, 2020), pp. 121--152.

\bibitem{lu_federated_2022}
M.~Y. Lu, R.~J. Chen, D.~Kong, J.~Lipkova, R.~Singh, D.~F. Williamson, T.~Y. Chen, and F.~Mahmood, \enquote{Federated learning for computational pathology on gigapixel whole slide images,} {\protect\JournalTitle{Medical Image Analysis}} \textbf{76}, 102298 (2022).

\bibitem{ogier_du_terrail_federated_2023}
J.~Ogier~du Terrail, A.~Leopold, C.~Joly, C.~Béguier, M.~Andreux, C.~Maussion, B.~Schmauch, E.~W. Tramel, E.~Bendjebbar, M.~Zaslavskiy, G.~Wainrib, M.~Milder, J.~Gervasoni, J.~Guerin, T.~Durand, A.~Livartowski, K.~Moutet, C.~Gautier, I.~Djafar, A.-L. Moisson, C.~Marini, M.~Galtier, F.~Balazard, R.~Dubois, J.~Moreira, A.~Simon, D.~Drubay, M.~Lacroix-Triki, C.~Franchet, G.~Bataillon, and P.-E. Heudel, \enquote{Federated learning for predicting histological response to neoadjuvant chemotherapy in triple-negative breast cancer,} {\protect\JournalTitle{Nature Medicine}} \textbf{29}, 135--146 (2023).

\bibitem{bouchama_fourier_2023}
L.~Bouchama, B.~Dorizzi, M.~Thellier, J.~Klossa, and Y.~Gottesman, \enquote{Fourier ptychographic microscopy image enhancement with bi-modal deep learning,} {\protect\JournalTitle{Biomedical Optics Express}} \textbf{14}, 3172 (2023).

\bibitem{jiang_solving_2018}
S.~Jiang, K.~Guo, J.~Liao, and G.~Zheng, \enquote{Solving {Fourier} ptychographic imaging problems via neural network modeling and {TensorFlow},} {\protect\JournalTitle{Biomedical Optics Express}} \textbf{9}, 3306 (2018).

\bibitem{claveau_digital_2020}
R.~Claveau, P.~Manescu, M.~Elmi, V.~Pawar, M.~Shaw, and D.~Fernandez-Reyes, \enquote{Digital refocusing and extended depth of field reconstruction in {Fourier} ptychographic microscopy,} {\protect\JournalTitle{Biomedical Optics Express}} \textbf{11}, 215 (2020).

\bibitem{zhou_analysis_2022}
H.~Zhou, C.~Shen, M.~Liang, and C.~Yang, \enquote{Analysis of postreconstruction digital refocusing in {Fourier} ptychographic microscopy,} {\protect\JournalTitle{Optical Engineering}} \textbf{61} (2022).

\bibitem{sitzmann2019siren}
V.~Sitzmann, J.~N. Martel, A.~W. Bergman, D.~B. Lindell, and G.~Wetzstein, \enquote{Implicit neural representations with periodic activation functions,} in \emph{Proc. NeurIPS,}  (2020).

\bibitem{park2019deepsdf}
J.~J. Park, P.~Florence, J.~Straub, R.~Newcombe, and S.~Lovegrove, \enquote{Deepsdf: Learning continuous signed distance functions for shape representation,} in \emph{Proceedings of the IEEE/CVF conference on computer vision and pattern recognition,}  (2019), pp. 165--174.

\bibitem{mildenhall2020nerf}
B.~Mildenhall, P.~P. Srinivasan, M.~Tancik, J.~T. Barron, R.~Ramamoorthi, and R.~Ng, \enquote{{NeRF: Representing Scenes as Neural Radiance Fields for View Synthesis},} {\protect\JournalTitle{European Conference on Computer Vision}}  (2020).

\bibitem{Pumarola_2021_CVPR}
A.~Pumarola, E.~Corona, G.~Pons-Moll, and F.~Moreno-Noguer, \enquote{D-nerf: Neural radiance fields for dynamic scenes,} in \emph{Proceedings of the IEEE/CVF Conference on Computer Vision and Pattern Recognition (CVPR),}  (2021), pp. 10318--10327.

\bibitem{feng2021signet}
B.~Y. Feng and A.~Varshney, \enquote{Signet: Efficient neural representation for light fields,} in \emph{Proceedings of the IEEE/CVF International Conference on Computer Vision,}  (2021), pp. 14224--14233.

\bibitem{feng2022prif}
B.~Y. Feng, Y.~Zhang, D.~Tang, R.~Du, and A.~Varshney, \enquote{Prif: Primary ray-based implicit function,} in \emph{European Conference on Computer Vision,}  (Springer, 2022), pp. 138--155.

\bibitem{feng2023neuws}
B.~Y. Feng, H.~Guo, M.~Xie, V.~Boominathan, M.~K. Sharma, A.~Veeraraghavan, and C.~A. Metzler, \enquote{Neuws: Neural wavefront shaping for guidestar-free imaging through static and dynamic scattering media,} {\protect\JournalTitle{Science Advances}} \textbf{9}, eadg4671 (2023).

\bibitem{chan2022efficient}
E.~R. Chan, C.~Z. Lin, M.~A. Chan, K.~Nagano, B.~Pan, S.~De~Mello, O.~Gallo, L.~J. Guibas, J.~Tremblay, S.~Khamis \emph{et~al.}, \enquote{Efficient geometry-aware 3d generative adversarial networks,} in \emph{Proceedings of the IEEE/CVF Conference on Computer Vision and Pattern Recognition,}  (2022), pp. 16123--16133.

\bibitem{zhu_dnf_2022}
H.~Zhu, Z.~Liu, Y.~Zhou, Z.~Ma, and X.~Cao, \enquote{{DNF}: diffractive neural field for lensless microscopic imaging,} {\protect\JournalTitle{Optics Express}} \textbf{30}, 18168 (2022).

\bibitem{liu_recovery_2022}
R.~Liu, Y.~Sun, J.~Zhu, L.~Tian, and U.~S. Kamilov, \enquote{Recovery of continuous 3d refractive index maps from discrete intensity-only measurements using neural fields,} {\protect\JournalTitle{Nature Machine Intelligence}} \textbf{4}, 781--791 (2022).

\bibitem{xie_diner_2022}
S.~Xie, H.~Zhu, Z.~Liu, Q.~Zhang, Y.~Zhou, X.~Cao, and Z.~Ma, \enquote{Diner: Disorder-invariant implicit neural representation,} in \emph{Proceedings of the IEEE/CVF Conference on Computer Vision and Pattern Recognition,}  (2023), pp. 1--10.

\bibitem{wang_local_2023}
H.~Wang, J.~Zhu, Y.~Li, Q.~Yang, and L.~Tian, \enquote{Local conditional neural fields for versatile and generalizable large-scale reconstructions in computational imaging,} {\protect\JournalTitle{arXiv}}  (2023). Publisher: {arXiv} Version Number: 2.

\bibitem{chen2022tensorf}
A.~Chen, Z.~Xu, A.~Geiger, J.~Yu, and H.~Su, \enquote{Tensorf: Tensorial radiance fields,} in \emph{European Conference on Computer Vision,}  (Springer, 2022), pp. 333--350.

\bibitem{fridovich2023k}
S.~Fridovich-Keil, G.~Meanti, F.~R. Warburg, B.~Recht, and A.~Kanazawa, \enquote{K-planes: Explicit radiance fields in space, time, and appearance,} in \emph{Proceedings of the IEEE/CVF Conference on Computer Vision and Pattern Recognition,}  (2023), pp. 12479--12488.

\bibitem{zuo_wide-field_2020}
C.~Zuo, J.~Sun, J.~Li, A.~Asundi, and Q.~Chen, \enquote{Wide-field high-resolution {3D} microscopy with {Fourier} ptychographic diffraction tomography,} {\protect\JournalTitle{Optics and Lasers in Engineering}} \textbf{128}, 106003 (2020).

\bibitem{yeh_experimental_2015}
L.-H. Yeh, J.~Dong, J.~Zhong, L.~Tian, M.~Chen, G.~Tang, M.~Soltanolkotabi, and L.~Waller, \enquote{Experimental robustness of {Fourier} ptychography phase retrieval algorithms,} {\protect\JournalTitle{Optics Express}} \textbf{23}, 33214 (2015).

\bibitem{tian_multiplexed_2014}
L.~Tian, X.~Li, K.~Ramchandran, and L.~Waller, \enquote{Multiplexed coded illumination for {Fourier} {Ptychography} with an {LED} array microscope,} {\protect\JournalTitle{Biomedical Optics Express}} \textbf{5}, 2376 (2014).

\bibitem{horstmeyer_solving_2015}
R.~Horstmeyer, R.~Y. Chen, X.~Ou, B.~Ames, J.~A. Tropp, and C.~Yang, \enquote{Solving ptychography with a convex relaxation,} {\protect\JournalTitle{New Journal of Physics}} \textbf{17}, 053044 (2015).

\bibitem{kingma2014adam}
D.~P. Kingma and J.~Ba, \enquote{Adam: A method for stochastic optimization,} {\protect\JournalTitle{arXiv preprint arXiv:1412.6980}}  (2014).

\bibitem{bian_autofocusing_2020}
Z.~Bian, C.~Guo, S.~Jiang, J.~Zhu, R.~Wang, P.~Song, Z.~Zhang, K.~Hoshino, and G.~Zheng, \enquote{Autofocusing technologies for whole slide imaging and automated microscopy,} {\protect\JournalTitle{Journal of Biophotonics}} \textbf{13} (2020).

\bibitem{konda_fourier_2020}
P.~C. Konda, L.~Loetgering, K.~C. Zhou, S.~Xu, A.~R. Harvey, and R.~Horstmeyer, \enquote{Fourier ptychography: current applications and future promises,} {\protect\JournalTitle{Optics Express}} \textbf{28}, 9603 (2020).

\end{thebibliography}

\end{document}